%% file: FSE_demo_main.tex
\documentclass[sigconf, screen]{acmart}
\AtBeginDocument{%
  }

\acmYear{2026}\copyrightyear{2026}
\setcopyright{cc}
\setcctype[4.0]{by}
\acmConference[FSE Companion '26]{34th ACM Joint European Software Engineering Conference and Symposium on the Foundations of Software Engineering}{July 5--9, 2026}{Montreal, QC, Canada}
\acmBooktitle{34th ACM Joint European Software Engineering Conference and Symposium on the Foundations of Software Engineering (FSE Companion '26), July 5--9, 2026, Montreal, QC, Canada}
\acmDOI{10.1145/3803437.3806427}
\acmISBN{979-8-4007-2636-1/26/07}
\usepackage{xspace}
\include{macro}
\begin{document}

%%
%% The "title" command has an optional parameter,
%% allowing the author to define a "short title" to be used in page headers.
\title{ \name: Python Specification Generation using Large Language Models}

%%
%% The "author" command and its associated commands are used to define
%% the authors and their affiliations.
%% Of note is the shared affiliation of the first two authors, and the
%% "authornote" and "authornotemark" commands
%% used to denote shared contribution to the research.
% \author{Ben Trovato}
% \authornote{Both authors contributed equally to this research.}
% \email{trovato@corporation.com}
% \orcid{1234-5678-9012}
% \author{G.K.M. Tobin}
% \authornotemark[1]
% \email{webmaster@marysville-ohio.com}
% \affiliation{%
%   \institution{Institute for Clarity in Documentation}
%   \city{Dublin}
%   \state{Ohio}
%   \country{USA}
% }

\author{Ragib Shahariar Ayon}
\affiliation{%
  \institution{Texas State University}
  \city{San Marcos}
  \state{TX}
  \country{USA}
}
\email{ipd21@txstate.edu}
\orcid{0009-0006-6372-5000}

\author{Shibbir Ahmed}
\affiliation{%
  \institution{Texas State University}
  \city{San Marcos}
  \state{TX}
  \country{USA}
}
\email{shibbir@txstate.edu}
\orcid{0000-0003-1183-883X}

% \author{Huifen Chan}
% \affiliation{%
%   \institution{Tsinghua University}
%   \city{Haidian Qu}
%   \state{Beijing Shi}
%   \country{China}}

% \author{Charles Palmer}
% \affiliation{%
%   \institution{Palmer Research Laboratories}
%   \city{San Antonio}
%   \state{Texas}
%   \country{USA}}
% \email{cpalmer@prl.com}

% \author{John Smith}
% \affiliation{%
%   \institution{The Th{\o}rv{\"a}ld Group}
%   \city{Hekla}
%   \country{Iceland}}
% \email{jsmith@affiliation.org}

% \author{Julius P. Kumquat}
% \affiliation{%
%   \institution{The Kumquat Consortium}
%   \city{New York}
%   \country{USA}}
% \email{jpkumquat@consortium.net}

%%
%% By default, the full list of authors will be used in the page
%% headers. Often, this list is too long, and will overlap
%% other information printed in the page headers. This command allows
%% the author to define a more concise list
%% of authors' names for this purpose.
%\renewcommand{\shortauthors}{Trovato et al.}

%%
%% The abstract is a short summary of the work to be presented in the
%% article.
% \begin{abstract}

% \end{abstract}

%%
%% The code below is generated by the tool at http://dl.acm.org/ccs.cfm.
%% Please copy and paste the code instead of the example below.
%%
\begin{CCSXML}
<ccs2012>
   <concept>
       <concept_id>10011007.10011074.10011099.10011692</concept_id>
       <concept_desc>Software and its engineering~Formal software verification</concept_desc>
       <concept_significance>500</concept_significance>
       </concept>
 </ccs2012>
\end{CCSXML}

\ccsdesc[500]{Software and its engineering~Formal software verification}
%%
%% Keywords. The author(s) should pick words that accurately describe
%% the work being presented. Separate the keywords with commas.
\keywords{Specification, Large Language Model, Python}
%% A "teaser" image appears between the author and affiliation
%% information and the body of the document, and typically spans the
%% page.
% \begin{teaserfigure}
%   \includegraphics[width=\textwidth]{sampleteaser}
%   \caption{Seattle Mariners at Spring Training, 2010.}
%   \Description{Enjoying the baseball game from the third-base
%   seats. Ichiro Suzuki preparing to bat.}
%   \label{fig:teaser}
% \end{teaserfigure}

%   \received{February 01 2026}
% \received[revised]{12 March 2009}
% \received[accepted]{5 June 2009}

%%
%% This command processes the author and affiliation and title
%% information and builds the first part of the formatted document.
\include{sections/abstract}
\maketitle

\input{sections/introduction}
\input{sections/relatedWork}
\input{sections/approach}

\input{sections/results}
\input{sections/usage}
\input{sections/limitations}
\input{sections/conclusion}
\input{sections/dataAvailiblity}

%%
%% The acknowledgments section is defined using the "acks" environment
%% (and NOT an unnumbered section). This ensures the proper
%% identification of the section in the article metadata, and the
%% consistent spelling of the heading.
% \begin{acks}
% GPT and the Claude models were used to conduct experiments. ChatGPT was used to improve the overall presentation.
% \end{acks}
\balance
%%
%% The next two lines define the bibliography style to be used, and
%% the bibliography file.
\bibliographystyle{ACM-Reference-Format}
\bibliography{referenceDemo}

%%
%% If your work has an appendix, this is the place to put it.
% \appendix
% \include{sections/appendix}
\end{document}

%% file: macro.tex
\usepackage[utf8]{inputenc}
\usepackage[table]{xcolor}
\usepackage{xspace}
\usepackage{multirow}
\usepackage{subfig} 
\usepackage{balance} 
\usepackage{rotating}
\usepackage{enumitem}
\usepackage{listings}
\usepackage{lscape}
\usepackage[ruled,vlined]{algorithm2e}
\usepackage{tabularx} % package not found in ACM list
\usepackage{wrapfig} % package not found in ACM list
\usepackage{xurl} % package not found in ACM list
\usepackage{flushend} % package not found in ACM list
\usepackage{tikz} % package not found in ACM list
\usepackage[normalem]{ulem} % package not found in ACM list
\usepackage[most]{tcolorbox} % package not found in ACM list
\usepackage{syntax} % package not found in ACM list

\usepackage{listings}
\newcolumntype{L}{>{\raggedright\arraybackslash}X} % left-aligned stretch
\newcolumntype{C}{>{\centering\arraybackslash}X}   % centered stretch
\newcolumntype{R}{>{\raggedleft\arraybackslash}X}  % right-aligned stretch

\definecolor{codegreen}{rgb}{0,0.6,0}
\definecolor{codegray}{rgb}{0.5,0.5,0.5}
\definecolor{codepurple}{rgb}{0.58,0,0.82}
\definecolor{backcolour}{rgb}{0.95,0.95,0.92}
\definecolor{medium-blue}{rgb}{0,0,1}
\definecolor{darkblue}{HTML}{003566}
\definecolor{lightgray}{gray}{0.97}

\hypersetup{
  colorlinks=true,
  linkcolor=black,
  urlcolor=medium-blue,
  citecolor=black
}

\lstdefinestyle{custompython}{
  backgroundcolor=\color{backcolour},
  commentstyle=\color{codegreen},
  keywordstyle=\color{magenta},
  numberstyle=\tiny\color{codegray},
  stringstyle=\color{codepurple},
  basicstyle=\scriptsize\ttfamily,
  breaklines=true,
  captionpos=b,
  numbers=left,
  numbersep=5pt,
  showstringspaces=false,
  tabsize=2
}
\newcommand{\name}{\textit{SpecPylot}\xspace}
\newcommand{\crosshair}{\texttt{CrossHair}\xspace}
\newcommand{\icontract}{\texttt{icontract}\xspace}

\newcounter{NumObservations}
\newcommand{\finding}[2][]{%
  \begin{tcolorbox}[
      enhanced,
      breakable,
      colback=gray!5,
      colframe=black,
      boxrule=0.5pt,
      arc=2pt,
      left=6pt,
      right=6pt,
      top=4pt,
      bottom=4pt,
      width=\linewidth
  ]
  \textbf{#1\ifx&#1&Finding~\arabic{NumObservations}\fi:}~#2
  \end{tcolorbox}
  \stepcounter{NumObservations}%
}

\usetikzlibrary{positioning,shapes.geometric,shapes.misc}

\newtcolorbox{examplebox}[1][]{
  enhanced,
  breakable,
  colback=gray!5,
  colframe=gray!50,
  boxrule=0.4pt,
  arc=2pt,
  left=6pt,
  right=6pt,
  top=4pt,
  bottom=4pt,
  width=\linewidth,
  fontupper=\small,
  title={#1},
  fonttitle=\bfseries,
  coltitle=black,
  halign title=center
}

%% file: sections/abstract.tex
\begin{abstract}
Automatically generating formal specifications could reduce the effort needed to improve program correctness, but in practice, this is still challenging. Many developers avoid writing contracts by hand, which limits the use of automated verification tools. Recent large language models (LLMs) can generate specifications from code, but these specifications often fail in terms of verification. The reason is syntax errors, overly strict constraints, or mismatches with program behavior. We present \name, a Python tool that synthesizes executable specifications for Python programs as \texttt{icontract} annotations and checks them using \crosshair's symbolic execution. The tool relies on LLMs to propose candidate contracts and uses \crosshair to validate them. When \crosshair finds a concrete counterexample, \name updates only the generated contracts and leaves the program itself untouched. In addition, the tool can produce coverage-driven \texttt{pytest} stubs and keep detailed execution artifacts that are useful during debugging. Overall, the evaluation indicates that \name is able to generate \crosshair-compatible contracts for most programs, but it also highlights the practical limits introduced by bounded symbolic exploration and differences in LLM behavior.
\end{abstract}

%% file: sections/introduction.tex
\section{Introduction}
\label{sec:intro}

Writing software specifications is still a slow and error-prone process, even for experienced developers. In many Python projects, explicit specifications such as preconditions, postconditions, or invariants are missing altogether, which limits the applicability of automated verification and testing techniques in practice. This gap between available analysis tools and real-world code has been widely observed in prior work on specification inference and contract-based verification~\cite{ernst2007daikon,cousot2013automatic}.

Recent LLMs appear promising in this space. Given source code, they can often articulate high-level behavioral constraints that resemble human-written specifications, as demonstrated in several recent studies on LLM-assisted specification and invariant generation~\cite{pei2023can,chakraborty2023ranking,wen2024enchanting}. However, specifications produced directly by LLMs are frequently problematic. They may be inconsistent, overly restrictive, or incorrect regarding the program’s actual behavior. Furthermore, this issue was also reported in prior work evaluating LLM-generated specifications under formal verification~\cite{specgen,formalbench}. Without validation, such contracts are difficult to reuse or trust.

We present \name, a contract synthesis and verification pipeline for Python that addresses this gap by combining LLM-based specification generation with solver-backed symbolic execution. \name first asks an LLM to produce executable contracts as icontract~\cite{ristin2017icontracthypothesis} annotations, and then relies on CrossHair to check whether those contracts are consistent with the program. If CrossHair finds a concrete counterexample, \name uses that information to trigger a refinement step. In that refinement step, the model is asked to revise the contracts while leaving the underlying code untouched. This design is inspired by recent verifier in the loop approaches for specification refinement, which were adapted to the Python ecosystem and executable contracts~\cite{specgen, zhang2022pythonbycontract}.

At a high level, \name treats specification generation and specification checking as two separate concerns. The LLM is responsible for proposing candidate semantic constraints. However, CrossHair acts as an execution based sanity check that can either confirm those constraints or expose concrete violations within a bounded search. By iteratively refining contracts using counterexamples, \name improves specification quality without requiring manual intervention from the human developer.

\name is implemented as a practical research tool rather than a mere prototype. It supports multiple LLM backends, configurable verification and refinement budgets, and optional coverage-driven test generation. Our evaluation shows that \name can generate checkable specifications for most programs in a small benchmark, while also highlighting the limitations imposed by bounded symbolic execution and the variability of LLM-generated contracts.

In summary, this work makes the following contributions.
\begin{itemize}
\item We designed and developed \name, a Python tool designed to generate executable specifications using LLMs, while also verifying them through symbolic execution. 
\item Our approach allows the use of concrete counterexamples from CrossHair to refine and correct any incorrect or overly strict contracts, without changing the original program. 
\item The system is built to be practical and user-friendly, supporting various LLM providers, offering user-defined limits for verification and refinement, and including optional test generation based on coverage analysis. 
\item We evaluated the tool on a small set of Python programs, providing insights into which types of specifications can be verified, how long the process takes, and the costs involved.
\end{itemize}

%% file: sections/relatedWork.tex
\section{Related Work/Research Gap}
\label{sec:related}
Earlier work on automatic specification generation relied on program analysis to extract properties from code. More recent approaches use learning-based techniques, including large language models. \name builds on this direction by focusing on how LLM-generated specifications can be made usable under execution-based verification. Before LLMs were widely used, most work on specification inference relied on static or dynamic analysis. Dynamic invariant detectors such as Daikon~\cite{ernst2007daikon} infer properties by observing execution traces, while static techniques based on constraint solving or abstract interpretation attempt to establish lightweight correctness guarantees~\cite{alshnakat2020constraint,cousot2013automatic}. Several hybrid approaches combine these ideas with symbolic execution or mutation strategies~\cite{molina2021evospex,chakraborty2023ranking}. Although these methods can be effective in controlled settings, they often depend on restricted specification grammars, limited language features, or substantial manual effort. In practice, this makes them difficult to apply to realistic Python programs.

More recently, researchers have explored using large language models to generate specifications directly from source code. This direction builds on the strong performance of LLMs in related tasks such as code generation~\cite{zeng2022extensive}, summarization~\cite{ahmed2022few}, and defect prediction~\cite{hou2024large}. Several studies report that LLMs can produce invariants or contracts that resemble those written by developers~\cite{pei2023can,chakraborty2023ranking,wen2024enchanting}. SpecGen~\cite{specgen}, for example, combines few-shot prompting with mutation-based heuristics to generate JML specifications that can be checked by a verifier. Other approaches filter model outputs using ranking or static analysis. A common limitation across this line of work is that specification generation is typically treated as a one-shot activity, using fixed prompts or a single model, with little support for diagnosing or repairing incorrect specifications. Alongside generation techniques, some efforts concentrate on evaluation rather than generation. FormalBench~\cite{formalbench} examines the robustness and consistency of LLM-generated specifications under different prompting strategies. Agent-based frameworks such as CodeVisionary~\cite{wang2025codevisionaryagentbasedframeworkevaluating} and RepoMasterEval~\cite{wu2024repomastereval} study collaborative reasoning in code generation more broadly. 
Recent work also explores verifier-in-the-loop approaches that use test oracles or formal feedback to refine generated annotations~\cite{faria2026automatic}. While these approaches show promising results in verification-aware languages such as Dafny, they target different ecosystems and rely on test assertions as primary specification oracles.

In the Python ecosystem, the Python-by-Contract dataset~\cite{zhang2022pythonbycontract} provides a collection of programs annotated with executable icontract specifications~\cite{ristin2017icontracthypothesis} and has been used to evaluate tools such as CrossHair~\cite{schanely2017crosshair}. These resources are valuable, but they assume specifications are written manually and therefore do not address the cost of producing contracts. Related work in other language ecosystems follows similar ideas under different assumptions. AutoReSpec~\cite{ayon2026autore} targets Java and JML, combining multiple LLMs with verifier feedback to improve robustness.  In contrast, SpecPylot focuses on Python, executable \texttt{icontract} specifications, and refinement guided by concrete counterexamples from symbolic execution. By closely coupling specification generation with execution-based validation, SpecPylot supports iterative refinement and enables analysis of both successful and failing cases in practice.

%% file: sections/approach.tex
\section{\name Approach}
\label{sec:approach}

\begin{figure}[!t]
    \centering
    \includegraphics[width=1\linewidth]{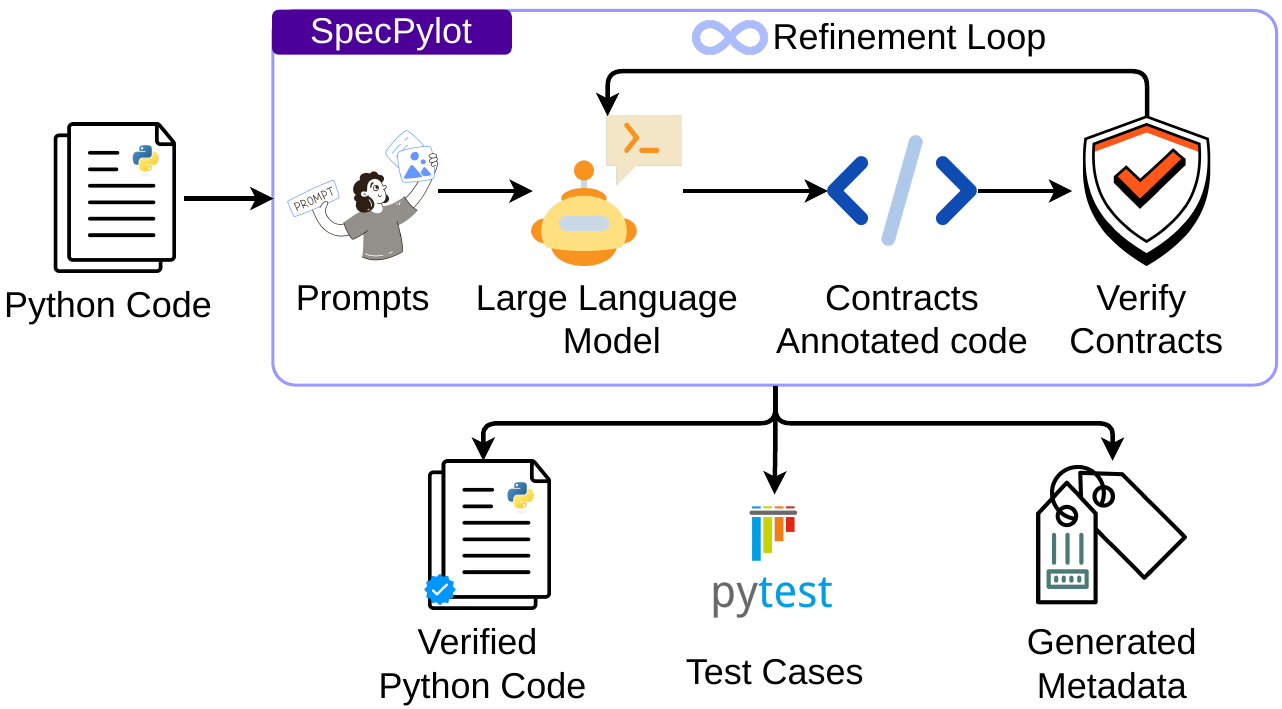}
    \caption{Overview of \name.}
    \label{fig:overview}
    \Description{Overview of SpecPylot, illustrating a pipeline where Python code is processed through prompts by a large language model to generate contract-annotated code, which is iteratively refined via a feedback loop and verified, producing verified code, test cases, and generated metadata}
\end{figure}
\name is a contract synthesis and verification pipeline for Python. Given a target program, it prompts an LLM to produce \icontract-annotated code and uses \crosshair to check whether the inferred contracts are consistent with the program behavior. Figure~\ref{fig:overview} summarizes the data flow of \name, including the verification and refinement loop and its termination conditions. Users can configure the LLM backend, refinement budgets, and \crosshair exploration limits (Section~\ref{sec:usage}).

\subsection{Contract generation}
\name renders an LLM request from prompt templates, and the target program, and the model returns a single, well-formed wrapper containing contract-annotated code. The pipeline validates the wrapper and basic syntactic well-formedness. If the output does not conform, \name retries a bounded number of times with explicit formatting repair instructions before proceeding.

Here we provide a concrete example of the contract produced by \name. Listing~\ref{lst:absolute-icontract} shows an \texttt{icontract}-annotated implementation of \texttt{Absolute.absolute} generated by \name. The example illustrates how \name expresses input assumptions using \texttt{@require} (e.g., restricting \texttt{num} to an \texttt{int}) and encodes postconditions using \texttt{@ensure} (e.g., ensuring the return value is a non-negative integer and matches the mathematical definition of absolute value for both non-negative and negative inputs). 

These examples are intended to make the generated contracts concrete and to facilitate reproduction and inspection of \name's outputs. The class-based formulation in the listing~\ref{lst:absolute-icontract} came from the dataset example rather than from any requirement of the approach. The approach itself is general and applies equally to ordinary top-level functions.

\begin{lstlisting}[caption={\texttt{icontract}-annotated implementation of \texttt{Absolute.absolute}.},label={lst:absolute-icontract}]
# icontract annotated code
from icontract import require, ensure
class Absolute:
    @require(lambda num: isinstance(num, int))
    @ensure(lambda result: isinstance(result, int))
    @ensure(lambda result: result >= 0)
    @ensure(lambda num, result: (num >= 0 and result == num) or (num < 0 and result == -num))
    def absolute(self, num: int) -> int:
        if 0 <= num:
            return num
        else:
            return -num
\end{lstlisting}
\vspace{-1.5em}

\subsection{Verification with \crosshair}
\name invokes \crosshair \texttt{check} with \icontract analysis on the annotated code.
The outcome is classified into three statuses used throughout the pipeline:
\texttt{PASSED} when no counterexample is found and the analysis is exhaustive under
the configured limits, \texttt{REFUTED} when \crosshair produces a counterexample for
at least one contract, and \texttt{INCONCLUSIVE} when \crosshair finds no counterexample
but cannot exhaustively explore the relevant paths under the limits.

\subsection{Budgeted counterexample-driven refinement}
Refinement is triggered only on \texttt{REFUTED} runs that satisfy a user-defined
budget policy. Concretely, a refutation qualifies for refinement only when a concrete counterexample is reported, the check run does not time out, and the refutation is produced within the refinement time budget. When refinement is allowed, \name builds a second LLM request containing (i) the current annotated code and (ii) a compact evidence bundle extracted from \crosshair (for example, the failing call and diagnostic context). The LLM revises only the contracts, and \name re-runs \crosshair on the revised code. This loop continues until the run becomes \texttt{PASSED}, becomes \texttt{INCONCLUSIVE}, or reaches the maximum refinement rounds.

\subsection{Optional test-case generation and artifacts}
When enabled, \name runs \crosshair \texttt{cover} on the final annotated code to emit coverage-driven examples as \textit{pytest} stubs. Each run stores the final annotated program and a structured summary of outcomes, including verification status, refutation evidence when present, and refinement traces. If logging is enabled, \name additionally records prompts, raw model outputs, intermediate code versions, and \crosshair stdout/stderr to support debugging and reproducible evaluation.

%% file: sections/results.tex
\section{Experimental Results and Evaluation}
\label{sec:results}
To evaluate \name, we randomly selected 16 SpecGenBench programs~\cite{specgen} across four control-flow categories (branched, single-loop, multi-path loop, and nested loop), translated them from Java to Python, and added four sequential Python programs, yielding 20 programs total across five types. We have used two best models, Claude Sonnet 4.5 and GPT-4o, inspired by prior research on JML specification generation~\cite{ayon2026autore, formalbench, specgen}. Experiments ran on an Apple M4 MacBook Pro (24GB RAM). \name is implemented in Python~3.12. We use this dataset to evaluate both the effectiveness and efficiency of \name.

\subsection{Effectiveness of \name}
\begin{figure}[!h]
  \centering
  \includegraphics[width=\linewidth,trim=0 0.6cm 0 0,clip]{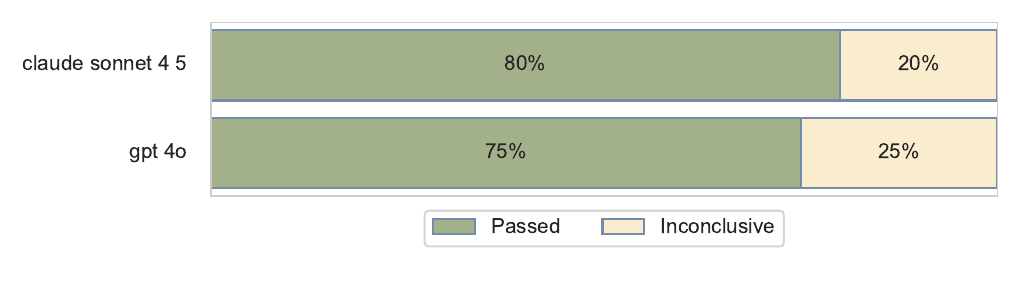}

  \caption{Pass and inconclusive rates of \name across 20 programs for Claude Sonnet 4.5 and GPT-4o.}
  \label{fig:rates}
  \Description{Pass and inconclusive rates of SpecPylot across 20 programs for Claude Sonnet 4.5 and GPT-4o. Claude Sonnet 4.5 passes 80\% of cases with 20\% inconclusive results, while GPT-4o passes 75\% with 25\% inconclusive results.}
\end{figure}

\begin{wrapfigure}{r}{0.54\linewidth}
% \vspace{-8pt}
\centering
\includegraphics[
  width=\linewidth,
  trim=0cm 0cm 0cm 0cm, clip]{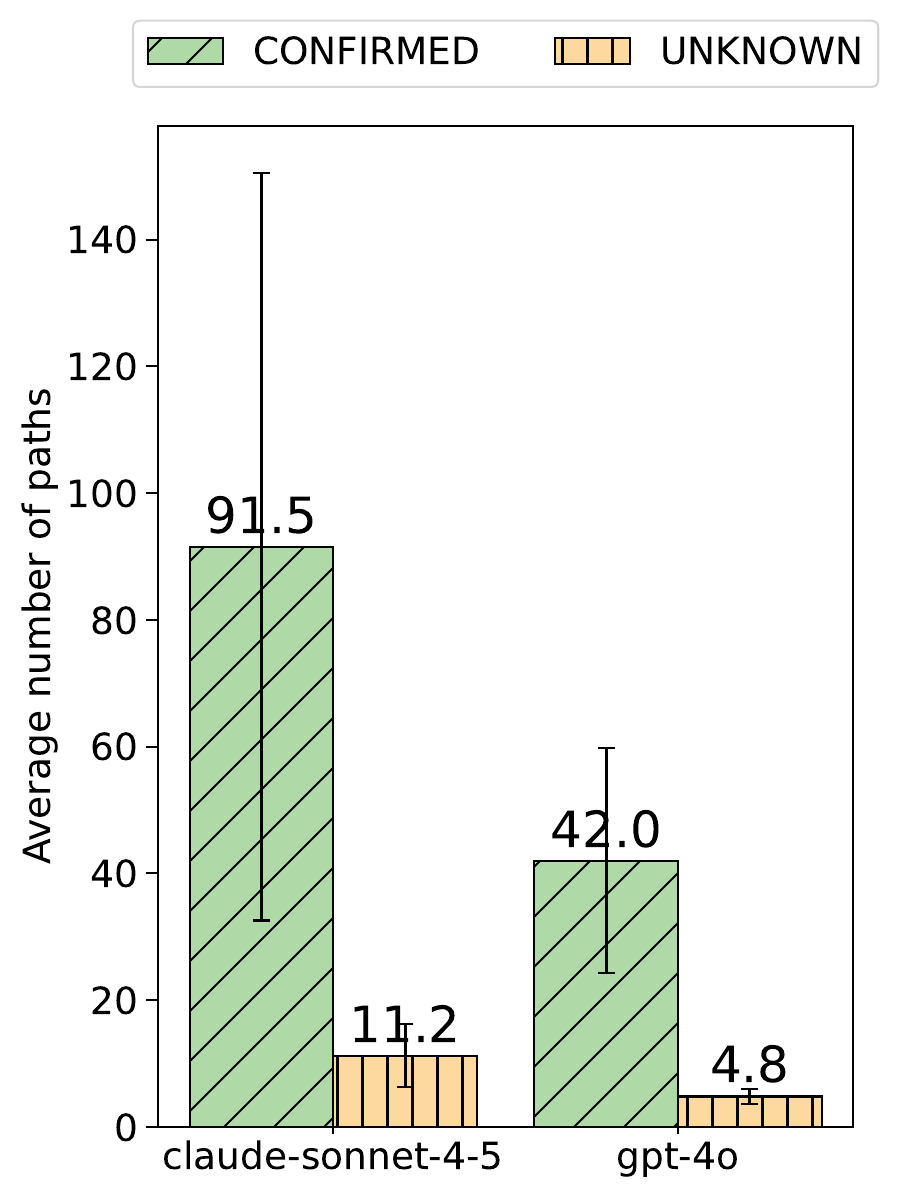}  
\vspace{-8pt}
\caption{Average \texttt{CONFIRMED} and \texttt{UNKNOWN} paths
explored by \crosshair for \texttt{INCONCLUSIVE} cases.}
\label{fig:avg-path-tested}
\Description{Bar chart showing the average number of paths explored by CrossHair for inconclusive cases. Claude Sonnet 4.5 explores an average of 91.5 confirmed paths and 11.2 unknown paths, while GPT-4o explores 42.0 confirmed paths and 4.8 unknown paths. Error bars indicate variability across programs.}
\end{wrapfigure}

To evaluate the effectiveness of \name, we report the \textit{pass rate} and \textit{inconclusive rate}, i.e., the fraction of programs labeled \texttt{PASSED} and \texttt{INCONCLUSIVE} out of the total dataset size. Figure~\ref{fig:rates} visualizes these rates as percentages for GPT-4o and Claude Sonnet 4.5. Using Anthropic's Claude Sonnet 4.5 with \name, we achieved a \textit{pass rate} of 80\% ($16/20$) and an \textit{inconclusive rate} of 20\% ($4/20$). Using OpenAI's GPT-4o with \name, we achieved a \textit{pass rate} of 75\% ($15/20$) and an \textit{inconclusive rate} of 25\% ($5/20$).

\texttt{INCONCLUSIVE} cases commonly arise for loop-heavy or highly branched code, where path explosion and timeout limits prevent exhaustive exploration; thus, the absence of a counterexample reflects an incomplete search rather than a proof. Figure~\ref{fig:avg-path-tested} reports the average number of paths explored for \texttt{INCONCLUSIVE} cases, and the error bars represent standard mean error.
For contracts generated by Claude Sonnet 4.5, \crosshair explored on average 91 \texttt{CONFIRMED} paths and 11.2 \texttt{UNKNOWN} paths; for GPT-4o, the averages were 42 \texttt{CONFIRMED} paths and 4.8 \texttt{UNKNOWN} paths. In our evaluation, the \texttt{INCONCLUSIVE} cases were 
\texttt{add\_loop}, \texttt{binary\_search}, \texttt{mysqrt}, and \texttt{digit\_root} (for both Claude Sonnet 4.5 and GPT-4o), and \texttt{calculator} (for GPT-4o). Across both models, these \texttt{INCONCLUSIVE} outcomes are associated with programs featuring multi-path loop control flow and/or nested loops, which expand the symbolic search space beyond what \crosshair can cover within the configured per-path and per-condition budgets, causing it to return an \texttt{UNKNOWN}-style outcome rather than \texttt{CONFIRMED} over all paths.

\subsection{Efficiency of \name}
We evaluate the efficiency of \name in terms of end-to-end runtime and API cost when paired with GPT-4o and Claude Sonnet 4.5. We conducted an efficiency comparison between Claude Sonnet 4.5 and GPT-4o when used with \name. Each point summarizes the average per-issue cost in two dimensions: the x-axis shows average end-to-end runtime (including contract generation, refinement, and \crosshair analysis), while the y-axis shows average token consumption (in thousands). Lower values on both axes indicate a more efficient configuration. In our evaluation, GPT-4o is faster on average, whereas Claude Sonnet 4.5 tends to use more tokens per issue. Both GPT-4o and Claude Sonnet 4.5 required refinement in only a small subset of tasks, averaging 1.2 total LLM iterations per run, indicating minimal reliance on the refinement loop. Refinement overlapped on \texttt{binary\_search.py} and \texttt{calculator.py}, with GPT-specific cases (\texttt{gcd.py, is_all_unique.py}) and a Claude-specific case (\texttt{divide.py}).
GPT consistently required a single refinement pass, whereas Claude required two passes for divide.py, resulting in slightly higher refinement effort in that case. Total runtime is primarily driven by closed-source LLM response latency and \crosshair's symbolic search, as well as test case generation. Overall, GPT-4o is faster and less expensive, while achieving a slightly lower \textit{pass rate} than Claude Sonnet 4.5. On average, Claude Sonnet 4.5 uses 5,480 tokens per issue, and GPT-4o uses 4,756 tokens per issue; both models converge quickly, requiring 1.02 refinement iterations on average. Across all 20 programs, the total API cost is \$1.09 for Claude Sonnet 4.5 and \$0.80 for GPT-4o.

%% file: sections/usage.tex
\section{Usage of \name}
\label{sec:usage}

\name is invoked from the command line. Users provide a target file and can set options that control provider selection, refinement behavior, \crosshair limits, and artifact logging. The pipeline stages are described in Section~\ref{sec:approach}. A typical invocation executing \name on \texttt{absolute.py} using GPT-4o, enabling 2 refinements and coverage-driven test generation as below:

\begin{lstlisting}[caption={},label={lst:run-specpylot}]
$ specpylot --target ./absolute.py --provider openai \
  --model gpt-4o --refine 2 --coverage
\end{lstlisting}

% \subsection{Configuration and controllable parameters}
% \label{subsec:configuration}
% %{\setlength{\parindent}{0pt}
% \name exposes a set of configuration parameters that let users trade off cost, runtime, and the amount of recorded artifacts.

% \paragraph{Provider and decoding.} Users select an LLM provider and model, and may tune decoding parameters (e.g., temperature). Credentials are supplied via environment variables or a local \texttt{.env} file loaded at startup.

% \paragraph{Retries and refinement.}
% Users configure the maximum number of retries for formatting/parsing failures, the number of refinement rounds, and an optional refinement budget. The budget restricts refinement to refutations produced within a user-specified time limit, thereby bounding additional LLM calls in difficult cases.

% \paragraph{\crosshair verification and test case generation.}
% \name performs verification using \texttt{crosshair check} and, when \texttt{--coverage} is enabled, generates coverage-driven \textit{pytest} stubs via \texttt{crosshair cover}. Users can separately configure timeouts and exploration controls for checking and coverage (overall timeout, per-condition timeout, per-path timeout, and maximum uninteresting iterations), which determine how thoroughly \crosshair explores execution paths.

% \paragraph{Outputs and logging.}
% Users specify output and log directories. When logging is enabled, \name records LLM prompts and responses, intermediate annotated code, and \crosshair stdout/stderr to support debugging and reproducible evaluation.
% }

\subsection{Configuration and controllable parameters}
\label{subsec:configuration}

\name provides a small set of configuration options that allow users to balance cost, runtime, and the amount of recorded artifacts. Users first select an LLM provider and model, and may adjust decoding parameters such as temperature. API credentials are supplied via environment variables or a local \texttt{.env} file, which is loaded at startup. The tool also exposes controls for retries and refinement. Users can configure the maximum number of retries for formatting or parsing failures, the number of refinement rounds, and an optional refinement budget. This budget limits refinement to refutations produced within a user-defined time window, which helps bound the number of additional LLM calls in difficult cases. Verification is performed using \texttt{crosshair check}. When coverage is enabled, \name additionally invokes \texttt{crosshair cover} to generate coverage-driven \textit{pytest} stubs. Timeouts and exploration limits can be configured independently for checking and coverage, including overall timeout, per-condition timeout, per-path timeout, and the maximum number of uninteresting iterations. These settings control how thoroughly \crosshair explores execution paths. Finally, users specify output and log directories. When logging is enabled, \name records LLM prompts and responses, intermediate annotated code, and \crosshair \texttt{stdout} and \texttt{stderr}, which supports debugging and reproducible evaluation.

%% file: sections/limitations.tex
\section{Limitations}
\label{sec:limitations}
This work has several limitations worth keeping in mind. When \crosshair reports a \texttt{PASSED} result, this should not be interpreted as a proof of correctness. It only means that no counterexample was found within the chosen exploration limits, and bugs may still exist. Similarly, \texttt{INCONCLUSIVE} results indicate that exploration stopped early, because of \texttt{UNKNOWN} paths, rather than that the contracts are correct. In our experience, programs with deep loops or complex branching quickly stress symbolic execution and lead to timeouts, which makes results sensitive to the configured budgets. The quality of results also depends on the typing information. \crosshair performs noticeably better when types are precise, while imprecise annotations can restrict the explored inputs or cause internal errors. Finally, the contracts generated by LLMs should be viewed as best-effort suggestions. They can be incomplete, overly strong, or simply unhelpful; results vary across models and configurations, and our conclusions are necessarily limited by the size of the evaluation.

%% file: sections/conclusion.tex
\section{Conclusion and Future Work}
\label{sec:conclusion}
In this work, we presented \name, a pipeline for automatically annotating Python programs with executable contracts and checking them using \crosshair’s solver-backed symbolic execution. We evaluated the tool with two widely used closed-source LLMs, GPT-4o and Claude Sonnet 4.5, and examined both how often the generated contracts could be verified and the time and cost of the end-to-end process. Across our benchmark, \name generated \crosshair-compatible contracts for most programs, but the results also highlight several practical limitations. In particular, bounded symbolic exploration and variability in LLM-generated specifications continue to affect verification outcomes. This is especially for programs with complex control flow. In the future, we can extend the tool to support PEP~316-style docstring contracts in addition to \texttt{icontract}. We also intend to explore LLM-agent workflows that operate at the repository level, using a broader project context to guide contract generation across multiple files and modules. We also plan to evaluate the approach on real-world contracts mined from recent GitHub repositories to assess generalization beyond LLM training data.

%% file: sections/dataAvailiblity.tex
\section{Data Availability and Tool Demo}
\label{sec:datapackage}
The reproduction package, evaluation results, and dataset are available in  Zenodo~\cite{ayon2026specpylotzenodo}, and we provide a tool demonstration in YouTube~\cite{SpecpylotDemo}. We hope these artifacts serve as a useful resource for future research in this area.